# Deep-learning-based acceleration of MRI for radiotherapy planning of pediatric patients with brain tumors

**Shahinur Alam** [1,*], **Jinsoo Uh** [3,*], **Alexander Dresner** [4], **Chia-ho Hua** [3], **and Khaled Khairy** [1,2,*]

[1] *Center for Bioimage Informatics, St. Jude Children's Research Hospital, Memphis , TN 38105, USA*
[2] *Department of Developmental Neurobiology, St. Jude Children's Research Hospital, Memphis , TN 38105, USA*
[3] *Department of Radiation Oncology, St. Jude Children's Research Hospital, Memphis , TN 38105, USA*
[4] *Philips Healthcare, Gainesville, FL, USA*

Correspondence*:
Shahinur Alam
Center for Bioimage Informatics, 262 Danny Thomas Place, St. Jude Children's Research Hospital, Memphis , TN 38105, USA
shahinur.alam@stjude.org

Khaled Khairy
Center for Bioimage Informatics, 262 Danny Thomas Place, St. Jude Children's Research Hospital, Memphis , TN 38105, USA, khaled.khairy@stjude.org

## ABSTRACT

Magnetic Resonance Imaging (MRI) is a pivotal non-invasive diagnostic and radiotherapy planning tool, offering detailed insights into the anatomy and pathological conditions of the human body, and in monitoring neurological, musculoskeletal, and oncological diseases. The relatively extensive scan time compared to other medical imaging modalities is stressful for patients, who must remain motionless in a prolonged imaging procedure that prioritizes reduction of imaging artifacts and an increase in data quality. This is particularly challenging for pediatric patients who require extra measures for managing voluntary motions such as anesthesia. Several computational approaches attempt to reduce the scan time (namely, fast MRI), which record fewer measurements and then digitally recover full information from this limited set in a post-acquisition reconstruction step. However, most fast MRI approaches have been developed with diagnostic MRI in mind, without addressing reconstruction challenges specific to scanning for radiation therapy (RT) planning. In this work, we developed a deep learning-based method called DeepMRIRec for MRI reconstruction from highly undersampled raw data acquired with RT-specific receiver coils. We evaluated our method against fully sampled k-space data of T1-weighted MR images acquired from 73 pediatric brains with tumors/surgical beds using loop and posterior coils (12 channels). We compared DeepMRIRec to several state-of-the-art deep learning-based reconstruction methods using our dataset with and without applying virtual compression of coil elements. DeepMRIRec reduced image acquisition time by a factor of four producing a structural similarity (SSIM) score of 0.96 $\pm$ 0.006, surpassing state-of-the-art methods and demonstrating its suitability for highly accelerated MRI scanning for pediatric RT planning.

## 1 INTRODUCTION

Magnetic Resonance Imaging (MRI) is an essential imaging modality in treatment planning for radiotherapy (RT) because of its superior soft tissue contrast defining target volumes and other anatomical structures. The absence of imaging-related ionizing radiation exposure in MRI is particularly beneficial for pediatric patients. However, one of the downsides of MRI is the relatively long scan time. This poses significant challenges in managing patient movement, impacts patient comfort and overall throughput. Younger patients may require anesthesia during the scan (typically patients under 8 years of age). Long scan times necessitate high anesthesia dosages, which raises the potential risk of neurocognitive consequences associated with accumulated anesthesia (Banerjee et al., 2019).

In conventional MRI, scan time is proportional to the number of phase encoding steps/lines captured in k-space. Accelerated MRI scanning (also known as fast MRI) can be achieved either by increasing the sampling rate or undersampling raw data in k-space, both of which produce unwanted artifacts. Examples of MRI sequences that increase sampling rate include turbo spin/gradient echo sequences and echo-planar imaging which require hardware enabling the desired strengths of radiofrequency/gradient pulses (Bernstein et al., 2004). Conventional undersampling methods include partial Fourier techniques (McGibney et al., 1993) and parallel imaging (Griswold et al., 2002; Pruessmann et al., 1999; Lustig et al., 2007; Lustig and Pauly, 2010), which utilize the conjugate symmetry in k-space and the spatial dependency of channel-wise data from individual coil elements, respectively, to reconstruct images containing sub-Nyquist artifacts (e.g., aliasing or Gibbs ringing). Such conventional fast MRI methods can be further advanced by low-rank modeling (Haldar, 2013) or compressed sensing (Lustig et al., 2007; Jaspan et al., 2015), which utilizes the sparsity of MR images in the transformed domain. However, sparse transformations are often limited in encoding complex image features, and the computational complexity results in lengthy reconstruction time. The major challenges of conventional MRI reconstruction methods are (Lebel, 2020):- 1) they force clinicians to sacrifice either image quality or spatial resolution, 2) reconstruction time is not well suited for clinical settings that demand low latency, 3) poor subsampling of k-space produces more artifacts, and 4) it is expensive as the number of coils increases. Hence, fast MRI acquisition and reconstruction are still an active research topic.

Recent advancement in Artificial Intelligence (AI) brought a breakthrough in object detection (Ahmed et al., 2018; Alam et al., 2020a; Alam, 2021; Alam et al., 2015, 2020b), behavioral and facial expression recognition (Anam et al., 2014), and super-resolution (Dong et al., 2014; Ledig et al., 2017) among others. Deep learning methodology, a branch of AI/ML (LeCun et al., 2015), and publicly available MRI datasets such as fastMRI(Zbontar et al., 2018) shifted the paradigm for MRI data reconstruction marking substantial progress in mitigating the temporal constraints of MRI scans. The latency of MRI reconstruction from undersampled k-spcae data is now shorter than ever with emerging deep learning based methods (Hammernik et al., 2018; Zhu et al., 2018; Hyun et al., 2018; Wu et al., 2023; Liu et al., 2021; Malkiel et al., 2019; Souza et al., 2019; Lebel, 2020). These computational models have been instrumental in fast imaging, reconstruction, and post-processing tasks, enabling the reduction of scan times with only small compromise in image quality. Various studies corroborate the effectiveness of these techniques in surpassing conventional approaches, showcasing the versatility and adaptability of machine learning and deep learning in optimizing acquisition strategies(Lebel, 2020).

Numerous machine learning and deep learning frameworks, such as Convolutional Neural Networks (CNN) and Recurrent Neural Networks (RNN), have been explored for their effectiveness in accelerating MRI scans. For example, Hammernik et al. (2018) developed a variational network for MRI reconstruction that preserves the natural appearance of MR images as well as pathologies; Zhu and colleagues (Zhu et al., 2018) presented a data driven approach called AUTOMAP (automated transform by manifold approximation) for mapping MRI sensor data to the image domain; Hyun and co-workers (Hyun et al., 2018) developed a U-net (Ronneberger et al., 2015) based model to reconstruct images from single channel MRI; Wu et al. (2023) used the Swin Transformer (Liu et al., 2021) and combined k-space consistency to improve MRI reconstructions; Lebel (2020) developed a deep CNN based pipeline called $AIR^{TM}$ Recon DL to remove truncation artifacts and increase sharpness in MRI images; Desai et al. (2023) proposed a self-supervised learning consistency training methods for MRI reconstructions.

However, the Deep learning-based fast MRI methods mentioned above have been developed with diagnostic images in mind and do not account for the unique challenges in MRI for RT planning. Unlike diagnostic imaging, RT-specific MRI on MR-RT simulators or MR-integrated linear accelerators often utilize receiver coil configurations with smaller numbers of coil elements and/or additional gaps from the body to accommodate immobilization devices or to minimize interference with radiation beams (Paulson et al., 2015; Raaymakers et al., 2017; Hua et al., 2018). This may result in compromised signal intensities in the channel-wise images (see Figure 1). Consequently, previously developed fast MRI methods may not be readily applicable as they are based on parallel MRI reconstructions utilizing redundancy across channel-wise images. Moreover, MRI reconstructions produced by conventional and emerging methods may not always be well suited for mission critical applications that require rich spatial information. Our goal, ultimately, is to develop automated tools that are sensitive enough for delineating tumors and for understanding biological processes such as the progression of pediatric cancers. This necessitates diagnostic images showing internal structures at high contrast; with easily discernible boundaries, and preservation of high frequency information while maintaining the lowest possible scan time. In addition, we need to ensure that radiation oncologists are not misled by a very plausible –but incorrect– reconstruction. Towards that end, we evaluated previously developed deep learning-based methods and propose an improved method that accounts for the RT coil configuration. We focused on cranial imaging in this study as the majority of our treatment cases involve brain tumors and brain images. Among the multitude of previous methods, we selected Hammernik et al. (2021) as our benchmark because it facilitates evaluation of multiple deep network models with various regularization networks and data consistency layers.

## 2 DATASET AND METHODS

### 2.1 Dataset Details

This study uses k-space raw data of T1-weighted MR images from 73 pediatric and young adult patients (aged 1-23 years) who were treated for common pediatric brain tumors at St. Jude Children's Research Hospital. The MR images were acquired on a 1.5T scanner (Ingenia, Philips Healthcare, Gainesville, FL, USA) for the purpose of planning radiation treatment. These scans utilized a receiver coil configuration accommodating immobilization devices such as a head mask and cushion, and which comprised flexible bilateral loop coils and the posterior coil embedded in the patient table. This coil configuration provided a total of 14 channels for the MR signals. We included the data from 12 channels; excluding 2 channels that gave weak signals and were inconsistently used. The MR images were acquired by a three dimensional turbo field echo sequence with the inversion recovery pre-pulse, where the frequency encoding was applied to anterior-posterior directions and phase encoding was applied to the other two directions. The repetition time (TR), echo time (TE), and inversion time (TI) were 2100ms, 3.5ms, and 1050ms, respectively. The scan time was typically 6 min and 25 s. To accommodate the two dimensional deep network, we mapped the original three dimensional data to two dimensions by performing a Fourier transform in the longitudinal dimension, and cropping out the superior or inferior slices of the brain. In addition, matrices of all images were adjusted to the same size of 256 × 192 by zero-padding to the periphery of k-space when smaller. The resulting field-of-view (FOV) in the left-right and anterior-posterior directions ranged from 159 to 183 mm and from 220 to 252 mm, respectively; in-plane resolutions were in the range of 0.83 – 0.95 mm and 0.86 – 0.98 mm, respectively. The slice thickness was 2 mm and the number of slices ranged from 56 to 69.

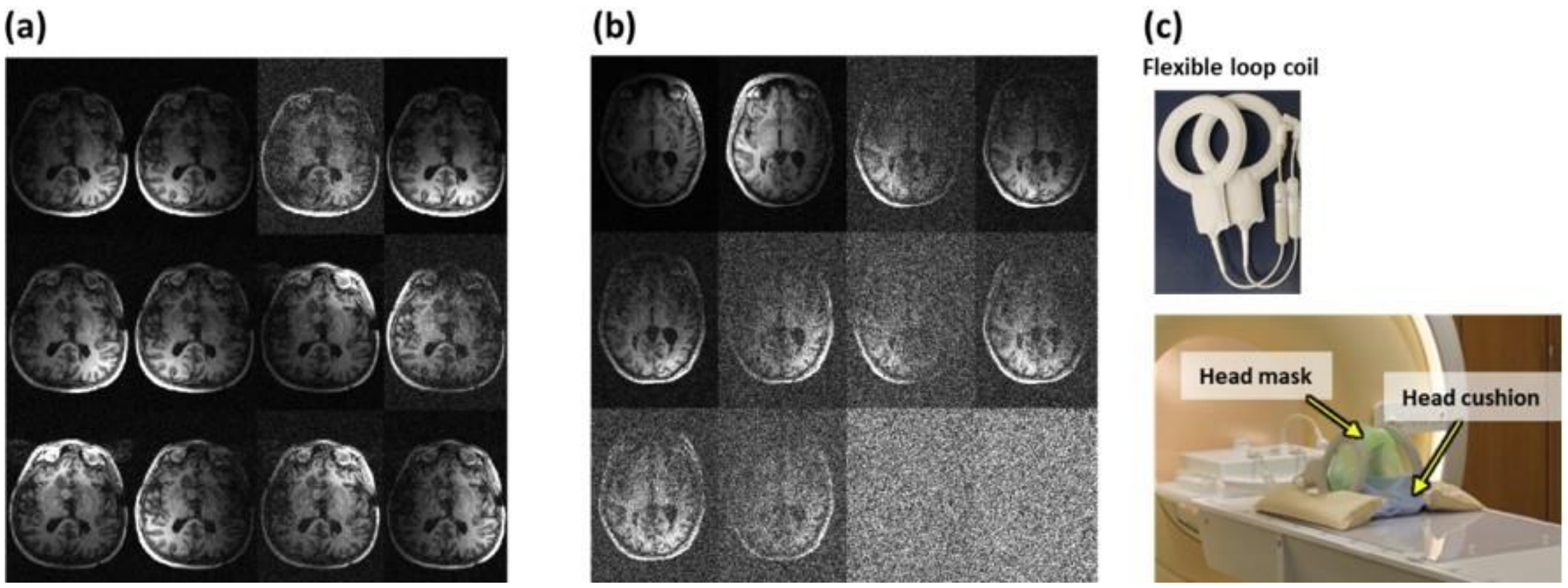

**Figure 1.** Comparison of channel-wise images from individual coil elements acquired by (a) a diagnostic 15 channel receiver coil versus (b) a loop coil-based configuration for RT imaging. (c) The loop coil-based coil configuration for cranial imaging is illustrated. In addition to the loop coils, coils embedded in the patient table are automatically engaged. Image intensities are often noisy and nonuniform across channels

## 2.2 Benchmark method

Reconstruction of MRI from undersampled raw data can be formulated (Equation 1) as an inverse problem (Knoll et al., 2020).

$$\hat{x} = argmin||Ax - y||^2 + \lambda R(x) \tag{1}$$

Here, x and y correspond to the vectorized image and raw data in k-space, respectively. The forward encoding operator A = MFS, is composed of an undersampling mask (M), Fourier transform (F), and sensitivity mapping (S). Since estimation of the true image x from the undersampled data is ill-posed, a regularization term R(x) providing apriori information with proper weighing $\lambda$ is required to avoid overfitting. In the evaluated benchmark method, a U-shaped network (UNET) (Ronneberger et al., 2015) and down-up network (DUNET) (Yu et al., 2019) compose the regularization term. Options for the gradient step of the data consistency terms included gradient descent (GD), proximal mapping (PM), and variable splitting (VS), which are detailed in the literature (Hammernik et al., 2021). A four-fold undersampling was simulated by applying a mask, that fully samples 10% of the central k-space and only 16.67% of the remaining 90% peripheral k-space, on the original fully sampled raw data. MR images reconstructed from the undersampled raw data are evaluated against images from the fully sampled data. Sensitivity maps are estimated by using the Berkeley Advanced Reconstruction Toolbox (BART;`https://github.com/mrirecon/bart`) which implements an eigenvalue method (namely, ESPIRiT) (Uecker et al., 2014) on the 10% fully-sampled central k-space data. Following the default hyperparameters, we used a learning rate of 0.0001, number of iterations 16, and initial $\lambda$ of 10. The network model was trained for 100 epochs with the RMSprop optimizer.

## 2.3 Summary of the DeepMRIRec workflow

The full workflow for MRI reconstructions is shown in Figure 2. In order to train our DeepMRIRec network, we first developed an undersampling mask (see section Undersampling k-Space) to find an optimal selection of sampling points and used this to undersample the full k-space data, simulating a four-fold scan speedup. Second, we applied the established Generalized Autocalibrating Partially Parallel Acquisitions (GRAPPA)[1] method to estimate missing values and reduce artifacts. Then, the GRAPPA-reconstructed k-space information was converted into image space using an Inverse Fourier Transformation (IFT). Finally, we developed a UNET-based model, which we trained on these GRAPPA-reconstructed images as an enhancer model to improve on the GRAPPA result. During inference, reconstruction of MRI images acquired at four-fold speed is performed following steps 3 to 6 in Figure 2. Comparison with the benchmark method is formulated as follows:

$$\hat{x} = UNET(argmin||Ax - y||^2 + \lambda||(G - I)Fx||^2) \tag{2}$$

The term in outer parenthesis indicates the GRAPPA-reconstructed image which satisfies data consistency (the first term) with the acquired data and calibration consistency (the second term) wherein the GRAPPA matrix G constrains the linear relationship between neighboring k-space points. A key difference from the benchmark method is that the deep network (UNET) encompasses the data consistency term as well, rather than composing only the regularization term. Subsequently, the input images to the network can be easily augmented by using conventional image transformations. By contrast, an augmentation of the k-space data would be required for the benchmark method, which is not readily feasible.

## 2.4 Undersampling K-Space

The primary goal of undersampling/subsampling is acceleration of image acquisition without loss of data quality. The byproduct of undersampling is an increase in aliasing and random image noise. Finding an efficient and optimal undersampling technique that reduces the aliasing effect received significant attentions in the recent literature (Razumov et al., 2023; Gaur and Grissom, 2015; Terpstra et al., 2020; Zhou et al., 2022). Although the majority of MRI is performed by acquiring k-space along a Cartesian trajectory, non-Cartesian (Wright et al., 2014) methods are also used (Zbontar et al., 2018; d'Arcy et al., 2002; Zhou et al., 2022). Non-Cartesian sampling has several benefits over Cartesian: 1) it is less sensitive to motion, 2) it produces better image contrast, and 3) it allows ultra-short echo times, among others. Unfortunately, it is more challenging to reconstruct images from non-Cartesian sampled k-space, and harder to implement in clinical settings. On the other hand, Cartesian sampling methods are straightforward and amenable to fast inverse Fourier transformation to reconstruct images from k-space. We explored various Cartesian undersampling methods (Zbontar et al., 2018) and proposed a new one that works best for both our dataset, as well as publicly available fastMRI data(Zbontar et al., 2018). We developed a mask (Figure 3) to retrospectively undersample full k-space data by a factor of 4. Our strategy is to give a higher sampling rate in the central k-space lines, despite the potential risk of compromising high-frequency information, to fully leverage the image information enriched towards the central k-space. The proposed 4-fold undersampling includes the fully sampled 10% central k-space as in the conventional scheme. However, the random sampling rates in the 90% peripheral regions was not spatially uniform. More specifically, the unilateral 45% peripheral region was divided into equally spaced 4 sections, and an 8:4:2:1 ratio of sampling rates were assigned from the proximal to distal sections (See Figure 3, four sections on each side of central k-space are separated by red dotted lines). The resultant acquisitions in these sections comprised 4%, 2%, 1%, and 0.5% of the entire k-space, making the sum of bilateral peripheral acquisitions 15% [i.e., $(4 + 2 + 1 + 0.5) \times 2 = 15$]. We added a random offset to pick k-space lines from left and right of the center, so that each data point carries distinctive information. The detailed algorithm and implementation are provided at `https://github.com/stjude/DeepMRIRec`.

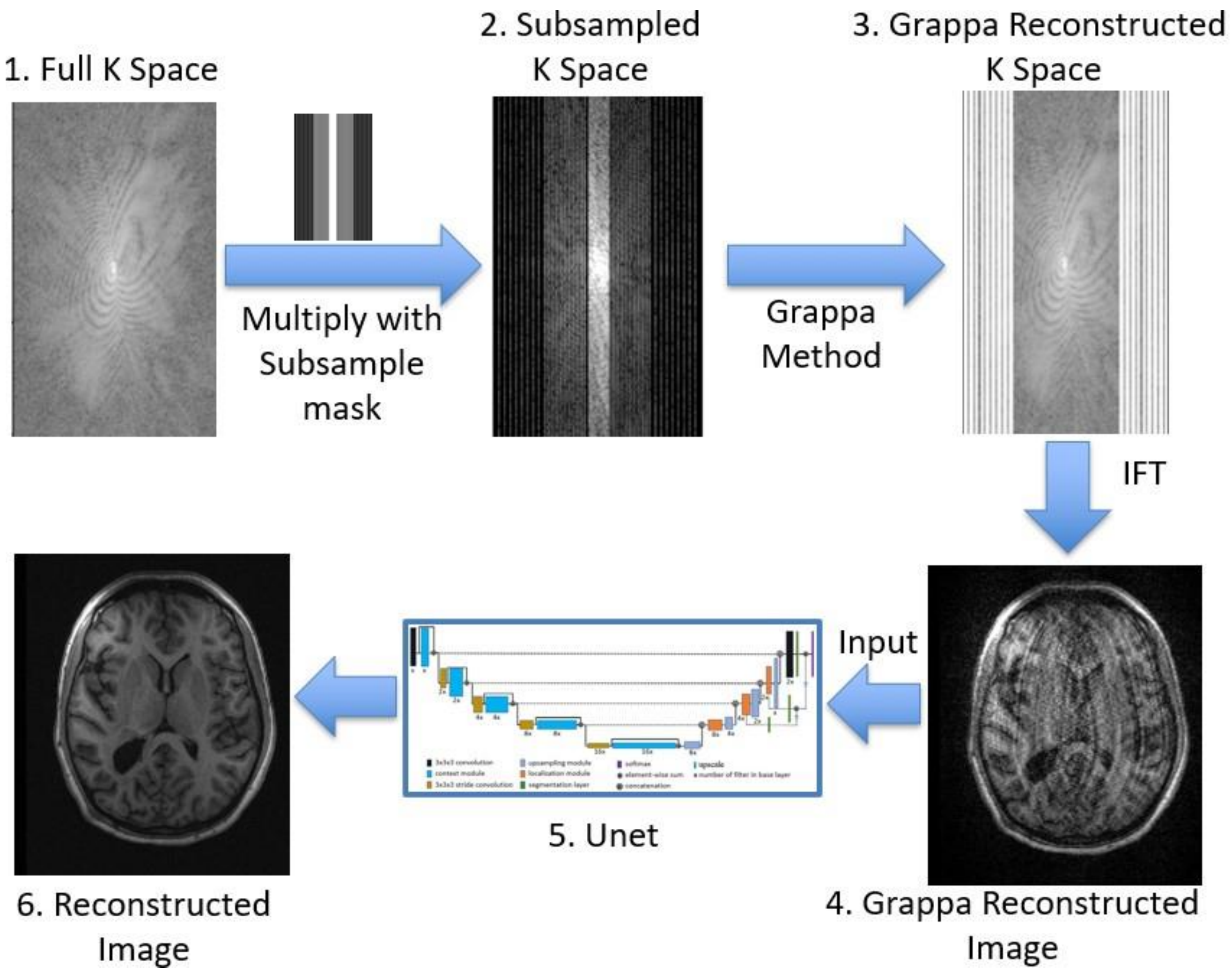

**Figure 2.** DeepMRIRec workflow: steps 1-6 for training and steps 3-6 for inference

## 2.5 Data Augmentation

Data augmentation is a technique to increase the diversity of a training dataset by applying realistic transformations. This technique reduces model overfitting (Shorten and Khoshgoftaar, 2019) and increases robustness against changes in tissue geometry, contrast, tissue density, field of view, orientation, and imaging conditions (Halevy et al., 2009; Sun et al., 2017).

We thoroughly evaluated various transformation models, image filters and associated parameters to generate 19 augmented images from each pair (reference and input). The reference images, $Y_{RSS}$ were obtained by applying Root Sum Square (RSS) on all coil-images obtained from full k-space (see Equation 3, K= full k-space, IFT=Inverse Fourier Transformation. The input undersampled images (I) were obtained using Equation 4, where M is the undersampling mask and G is the GRAPPA operation. Table 1 shows the list of transformation models and parameters used. A sample outcome of data augmentation with an elastic deformation model is shown in **Figure 4**.

$$Y_{RSS} = \sqrt{(IFT(K)} \qquad (3)$$

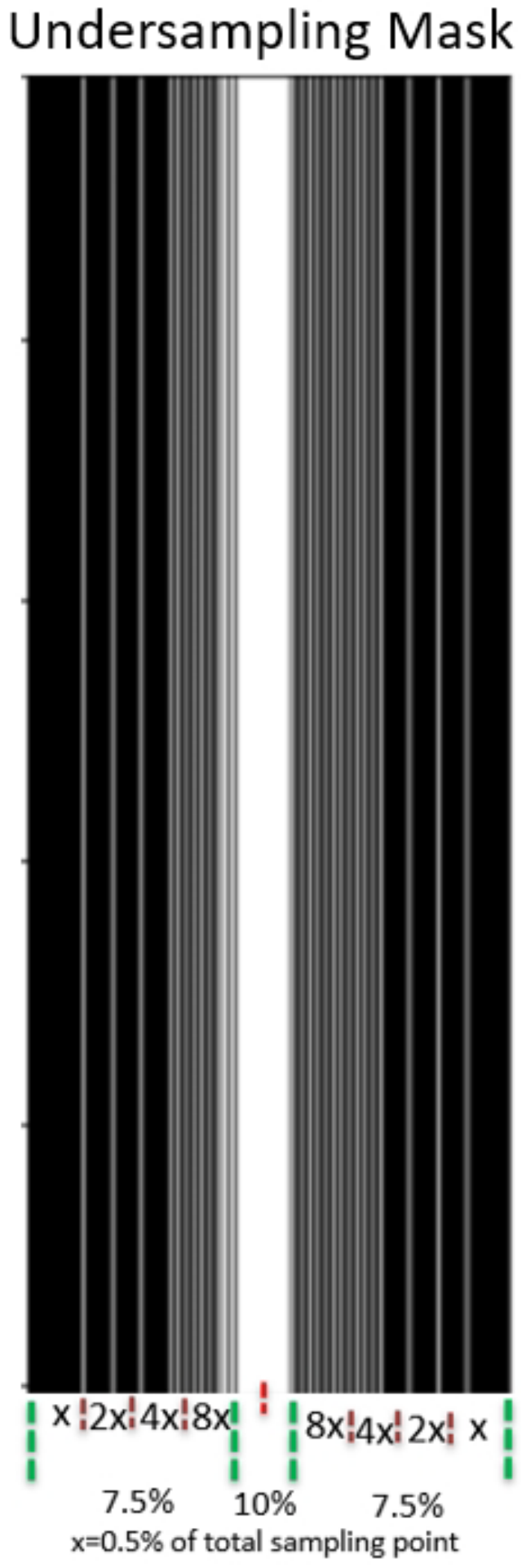

**Figure 3.** This mask subsamples full k-space by a factor of 4. 10% of the data points are chosen from the central region and 4%, 2%,1% and 0.5% from peripheral area.

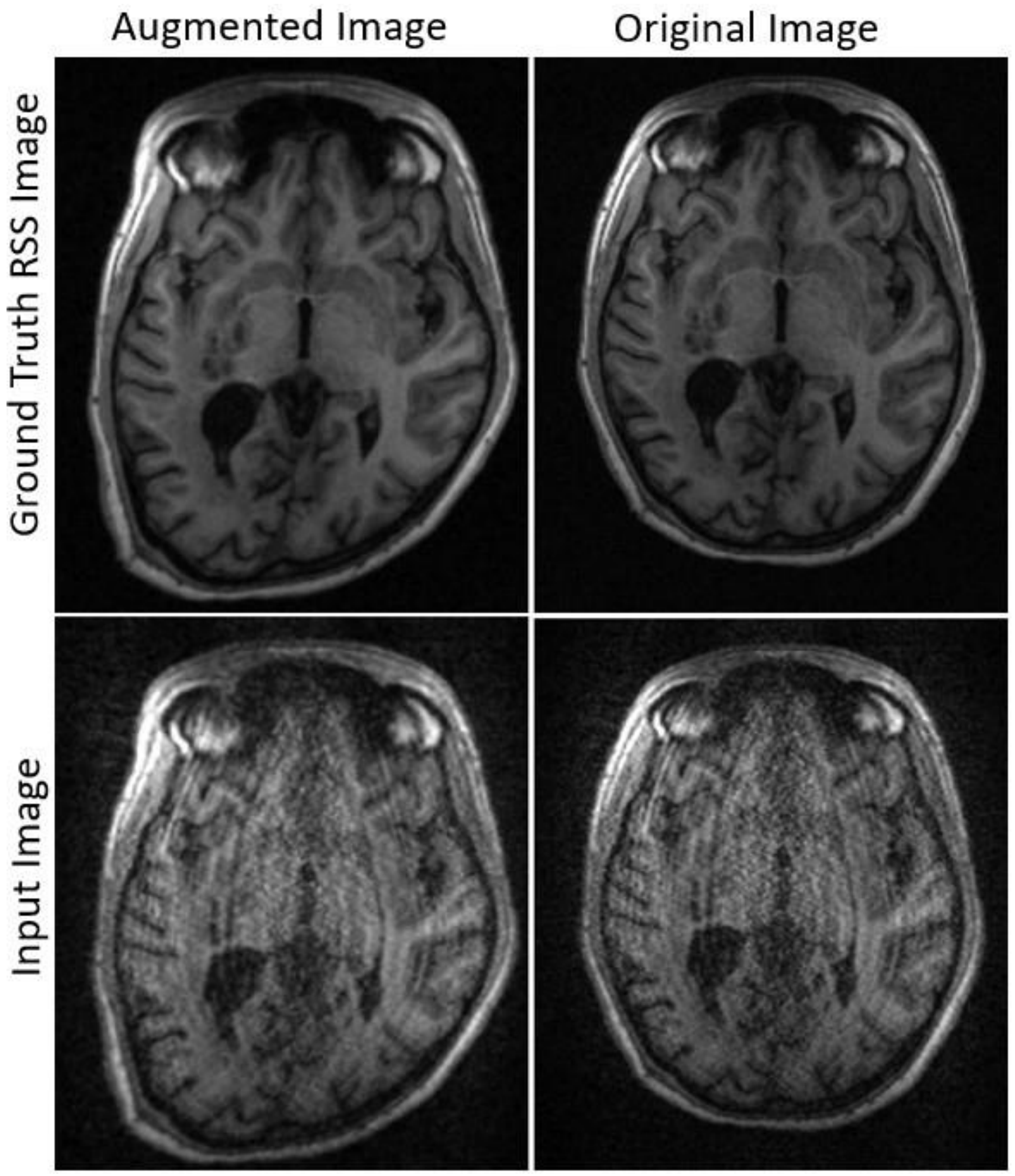

**Figure 4.** Data augmentation: Left column: augmented images. Right column: original images (input and ground truth pairs). Elastic deformation for a sample image is shown.

$$I = IFT(G(K * M)) \quad (4)$$

**Table 1.** Methods/Transformation models used for data augmentation. Values of transformation parameters are randomly selected from the range presented within the square brackets.

| Method's Name | Parameter Value/Ranges | Augmentation Nature |
|---|---|---|
| Horizontal Flip | | Produce horizontally flipped images |
| Dropout | [0.01, 0.05 ] | Creates images by dropping 1 to 5% voxels |
| Additive Gaussian Noise | scale [0.0, 12.75] | Creates images by adding noise sampled from Gaussian distributions |
| GaussianBlur | sigma [0.8, 1.5 ] | Creates smoothed images |
| Piecewise Affine | Scale[0.01, 0.07] | Creates images applying an affine transformation to a local grid |
| Elastic Transformation | alpha[1] [2.5, 50], sigma[2] [1,11] | Creates images by moving voxels locally |
| Affine Transformation | rotation along Z axis [-20°,20°] scale[0.7 ,1.5] isotropic Translation [-0.01%, 0.01%] | Creates images by applying an affine transformation |
| Rotation | along Y axis [-30°,30°] | Creates images by rotating around the Y axis |

## 2.6 RT-coil Compression

DeepMRIRec requires, on average, 120 ms to reconstruct an image from a single RT-coil, including GRAPPA which requires an average of 115 ms. The reconstruction time of an image by DeepMRIRec depends on the specific computational resources and the number of channels. Given the highly imbalanced signal-to-noise ratios across various channels, excluding those with weak signal strength might be beneficial for reconstruction efficiency, as well as for stabilizing model performance. However, the unique region-specific information provided by the excluded channels would be lost. Therefore, we investigated the potential of a virtual coil compression technique proposed by Zhang et al. (2013) to enhance the efficiency of image reconstruction with preserved or even improved quality of reconstructed images. In this technique, noisy images are incorporated into high-signal images via an optimized mapping in k-space, thereby the number of coil elements can be virtually reduced without excluding the region-specific information. In order to find an optimal number of virtual coils, we compressed the original 12 coil images (Figure 5.a) into 2, 3, and 4 virtual coil images (Figure 5.b), each of which required a separate training of DeepMRIRec.

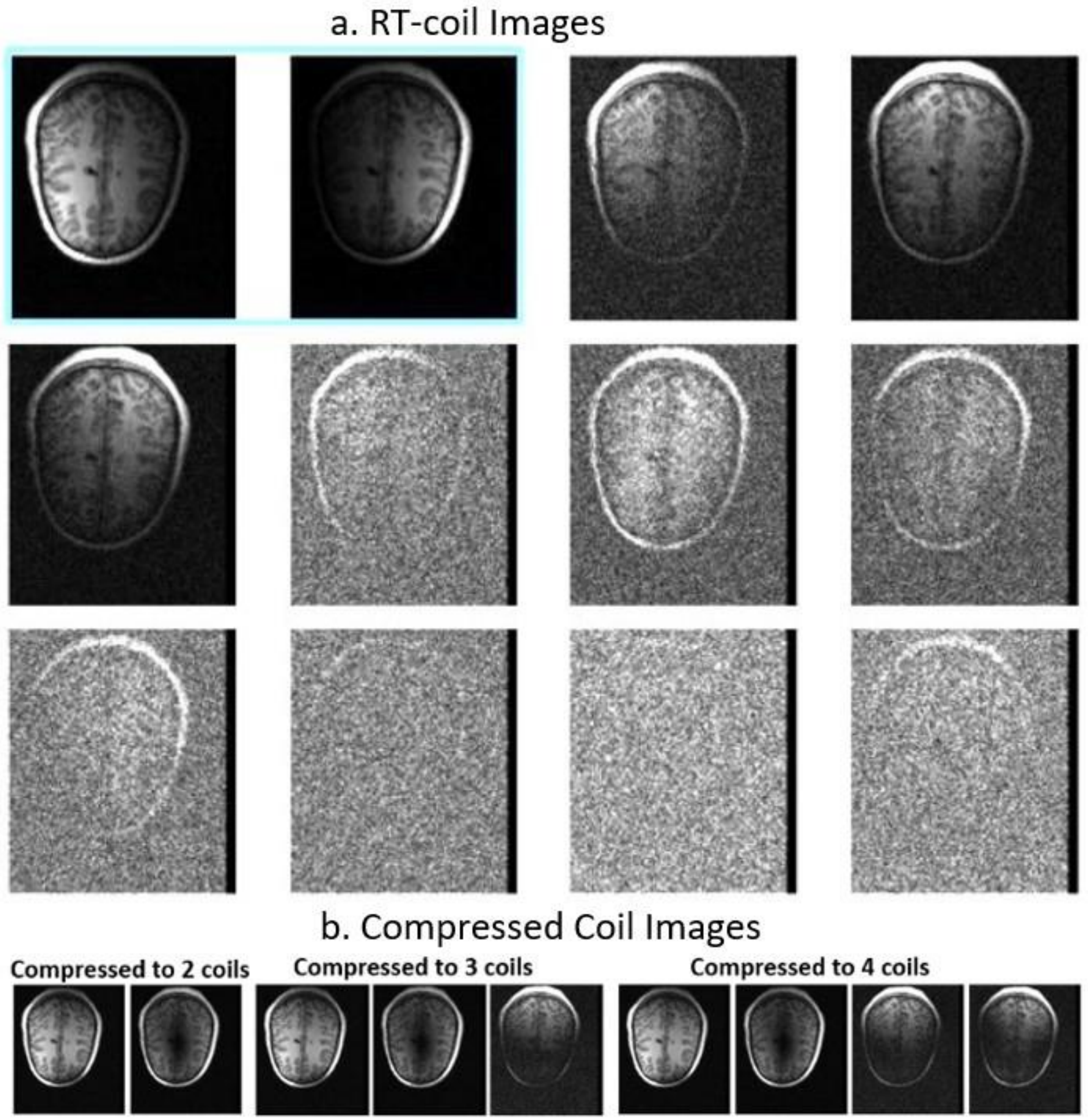

**Figure 5.** a. RT coil-image for a sample MRI. The signal-to-noise ratio among RT coils are not balanced across MRI volumes; b. Compressed coils

## 2.7 Deep Learning Model

### *2.7.0.1 Network Architecture*

We designed the DeepMRIRec network architecture (see **Figure 6**) following concepts from the well-known U-net (Ronneberger et al., 2015), ResNet (He et al., 2016) and DeepBrainIPP (Alam et al., 2022). DeepMRIRec consists of contracting/encoder (left) and expansive/decoder (right) paths. The contracting path extracts low-level features from input images and compresses/encodes them into high-level abstractions. The decoder recombines low-level features with higher-level abstractions through skip connections and performs precise localization. It also has a residual unit (He et al., 2016) to prevent vanishing gradients which is inherent in very deep networks. On the encoder side, at each layer spatial resolution is reduced by MaxPooling and the number of filters is increased. On the decoder side, feature dimension is reduced and higher-level image representations are upsampled to match ground truth images. The encoder and decoder contain a convolution block in each layer, which is comprised of two two-dimensional convolution operations (with a 3x3 kernel) followed by BatchNormalization, PreRelu and Dropout. The BatchNormalization is used to prevent co-variant shift, speed up training, and ease weights initialization. Unlike contemporary approaches, we considered depth, learning rate, dropout rate, and number of filters in the base layer as hyperparameters. We used a Bayesian Optimizer (Snoek et al., 2012) and Keras Tuner (O'Malley et al., 2019) packages to find optimal values of hyperparameters.

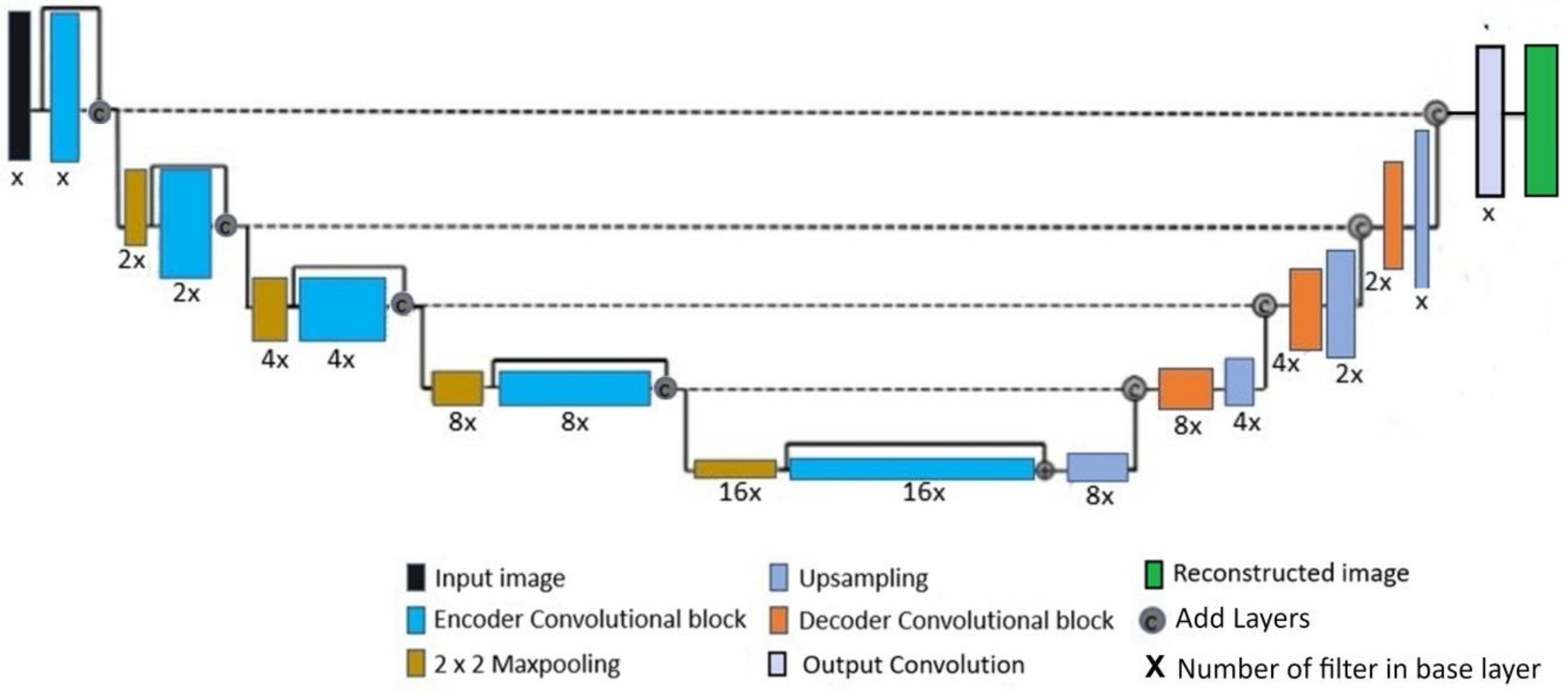

**Figure 6.** Network Architecture: We designed the DeepMRIRec network architecture by following paradigms from Unet and ResNet. We considered depth and number of filters in the base layer as hyperparameters, which were tuned by Bayesian optimization and grid search.

#### *2.7.0.2 Model Training*

DeepMRIRec was trained on 85348 2D image pairs of dimension 256x192 for 250 epochs, with an early stopping if the validation loss did not improve, using the Adam optimizer (Kingma and Ba, 2014) on an NVIDIA A100 80GB GPU system. We used a batch size of 128 and an initial learning rate of 0.0003 with reduce on Plateau. 80% of the dataset was used for training and the remaining 20% for validation. We performed channel/coil-wise range normalization of our data. The loss function used to train DeepMRIRec is shown in equation 5, where $\alpha = 0.0001$, $\beta = 1.0$, and $\gamma = 100$ are regularization parameters obtained via a grid search. The performance of our model was enhanced by: (1) data augmentation (from 4492 to 85348 2D images), (2) non-uniform undersampling in the peripheral k-space, (3) Bayesian optimization of hyperparameters, and (4) incorporation of weighted content loss along with L1 and SSIM loss (see loss function in Equation 5). The conventional loss function such as L1/NMSE is very sensitive to changes in contrast and luminous intensity. Therefore, we added content loss to minimize the effect of changes in photometric properties. The content loss was obtained from feature maps ($F_{GT}, F_S$) extracted from reference and predicted images using a VGG19 (Simonyan and Zisserman, 2014) pretrained model (see Equations 7 and 8). Figure 7.a shows high level features extracted using VGG19 for various convolutional layers. The content loss was calculated with layer-wise mean square error (MSE) between feature maps and weighted based on the depth of the convolutional block (see Equation 6). The optimal weights, ($\theta_i$), for convolutional blocks (i=1, 2, 3 and 4) found from a comprehensive grid search are 0.001, 0.01, 2, and 4 respectively. The optimal values for hyper parameters (learning rate, number of base filters of Unet, dropout rate, and depth of Unet) were obtained from Bayesian optimizers and determined to be 0.001, 32, 0.05, and 5 respectively.

$$Loss = \alpha * ContentLoss + \beta * L1Loss + \gamma * SSIMLoss \quad (5)$$

$$ContentLoss = \sum_{i=1}^{4} \theta_i * \frac{(F_{GT}(i) - F_S(i))^2}{L(F_{GT}(i))} \quad (6)$$

$$F_{GT} = VGG19(Y_{RSS}) \quad (7)$$

$$F_S = VGG19(DeepMRIRec(I)) \quad (8)$$

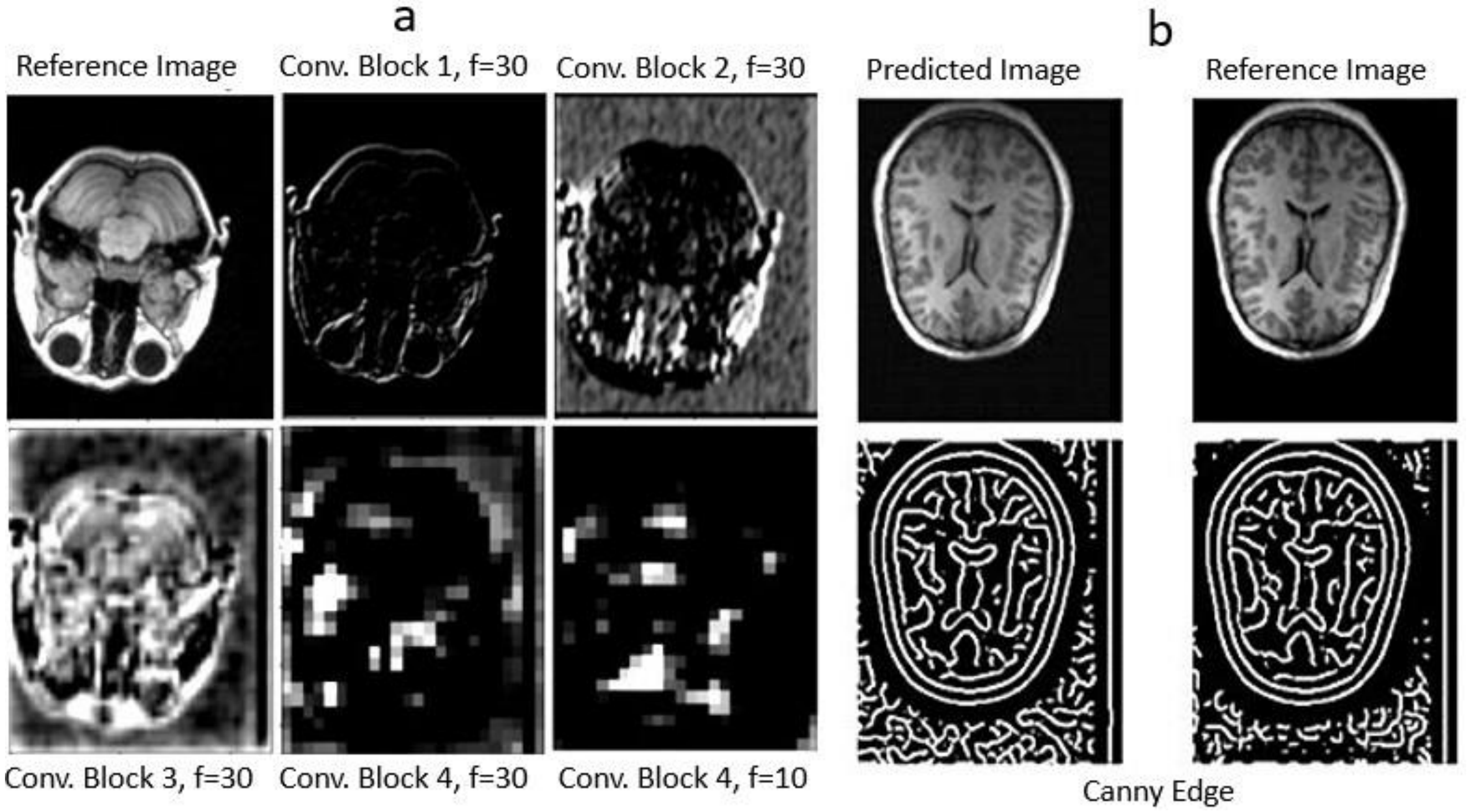

**Figure 7.** a) Extracted features using VGG19 to calculate content loss. f: index of activation/feature map b) high frequency components extracted from predicted and reference images using Canny edge detector to demonstrate how internal structures are aligned.

#### 2.7.0.3 Model Evaluation Metrics

The proper evaluation metric for MRI reconstruction is still a research question. Global metrics, such as SSIM, PSNR and NMSE, do not capture the full spectrum of information to assess the quality of reconstructed images (Nilsson and Akenine-Möller, 2020; Zbontar et al., 2018). PSNR and NMSE are sensitive to photometric properties such as brightness, contrast, hue and saturation, and do not reflect the true noise level present in images (Sara et al., 2019). SSIM compares luminance, contrast, and structural information between two images. However, SSIM has several limitations (Pambrun and Noumeir, 2015): 1) distortions at edges are underestimated, 2) insensitivity to high intensity regions, and 3) instability in regions with low variance. To address issues of PSNR, NMSE and SSIM, we also included the Dice Coefficient, and Hausdorff Distance (HD) on high frequency components (HFC) that ensure internal structures are properly reconstructed. HFC were extracted as a binary map using the Canny Edge Detector with a sigma of 5. Figure 7. b shows high frequency components extracted from predicted and reference images. The Dice coefficient is calculated from volumetric similarity using equation 9. To quantify the difference in edges we measured a Hausdorff Distance using equation 10. R and Y represent pixels in the binary map extracted from reconstructed and reference (full sampled) images. d(r,y) is the Euclidean distance between to binary maps.

$$Dice = 2(|R \cap Y|)/(|R| + |Y|) \tag{9}$$

$$\begin{aligned} Hausdorff &= max(h(R, Y), h(Y, R)) \\ h(R, Y) &= max(min(d(r, y))) \end{aligned} \tag{10}$$

## 3 RESULTS

The quantitative evaluation of DeepMRIRec was performed on 20% of our dataset. We present MRI reconstruction outcomes of our model along with several state-of-the-art deep learning-based models in Figure 8. Our model reconstruction produces SSIM of 0.96 $\pm$ 0.006, PSNR of 28, and Dice Coefficient, and HD on HFC are 0.66, and 72 respectively when applied to our dataset. We also trained and validated our model on a subset of publicly availabile fastMRI datasets (Zbontar et al., 2018) and it produced a SSIM of 0.98 whereas the reported best model output to date is 0.96 (Muckley et al., 2021). To test the effect of data augmentation and content loss we find that both (implemented together) increased our model score (SSIM) by 4.2%. Figure 9 a-b shows outcomes of our model with and without data augmentation. We also tested the effect of our proposed undersampling strategy, and find an increase in performance (SSIM) by 4.0% when compared to conventional methods (see Figure 9 c-d). The coil compression technique improved our reconstruction score (SSIM) by 1.0% over using the full set of twelve uncompressed coils. Finally, we investigated reconstruction outcomes when using two, three, and four compressed coils (Table 2). Reconstruction outcomes were found to degrade as we increase the number of virtual coils (data not shown) because the third, fourth and successive virtual coils contain more noise (see Figure 5.b).

## 4 DISCUSSION

Qualitative evaluation by two radiation oncologists, and the quantitative measures reported above, demonstrate the utility of DeepMRIRec. As shown in Figure 8, DeepMRIRec produced significantly better reconstruction outcomes, when compared to other MRI reconstruction models. This is the case when applied to our dataset, as well as when applied to publicly available fastMRI data. The proposed undersampling strategy (mask), incorporation of GRAPPA to pre-fill the undersampled k-space, data augmentation, simplified network architecture with Bayesian optimization, and using a customized loss function to train the model are all factors that collectively permit us to achieve this performance. Although GRAPPA adds significant computational overhead in our pipeline, it improves performance significantly (by 9.0%). It is important to keep in mind that the time consumption by GRAPPA increases linearly with the number of RT-coils. The coil compression technique played a significant practical role in decreasing the effective number of coils, and consequently saved the time needed to apply GRAPPA. Moreover, this decreased effective number of coils reduced training and inference times, as well as memory requirements for computation because the input dimension of the network is smaller (two channels instead of twelve).

The results also show that our model, trained on only loop coils, reconstructed images more accurately than when trained on all coils and that coil-compression not only speeds up MRI reconstructions, but also recovers the internal details of images more accurately. This is likely due to the noise-suppression and high frequency component retention inherent in the compression algorithm itself. Without reliance on coil-compression, network filters could not capture meaningful information in the earlier layers from noisy images. Instead, the low SNR lead to ambiguities in the higher layers.

Finally, our evaluation of selected models, including DeepMRIRec, revealed that they produce slightly smoothened images when compared to the fully sampled ones. One possible reason is that the undersampling mask contains more information from the center of the k-space image and picks less information from peripheral regions, which tend to contain high frequency information. Better future strategies for undersampling can mitigate this effect and increase reconstruction quality further.

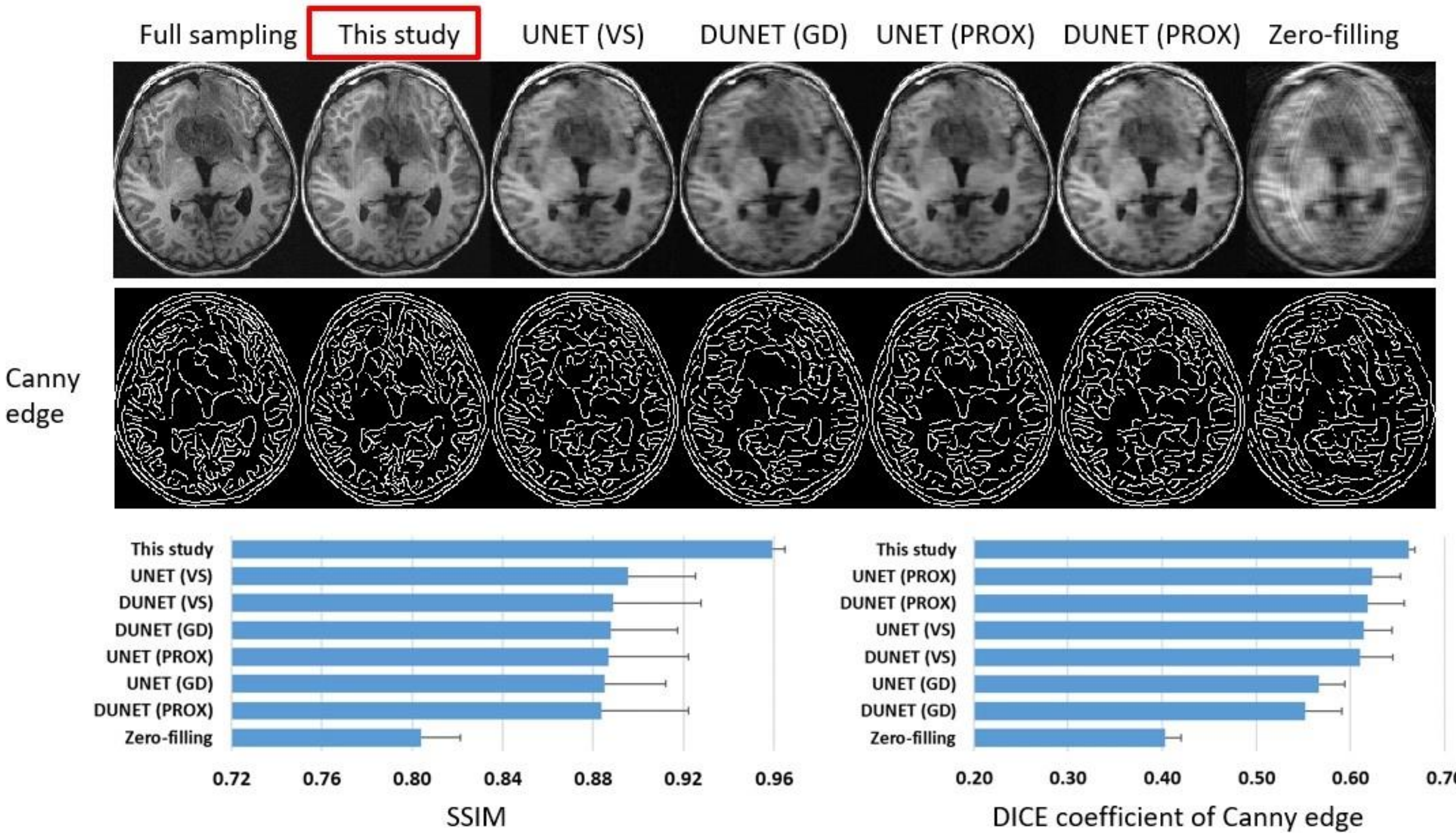

**Figure 8.** Comparison of reconstructed images using DeepMRIRec versus state-of-the-art methods

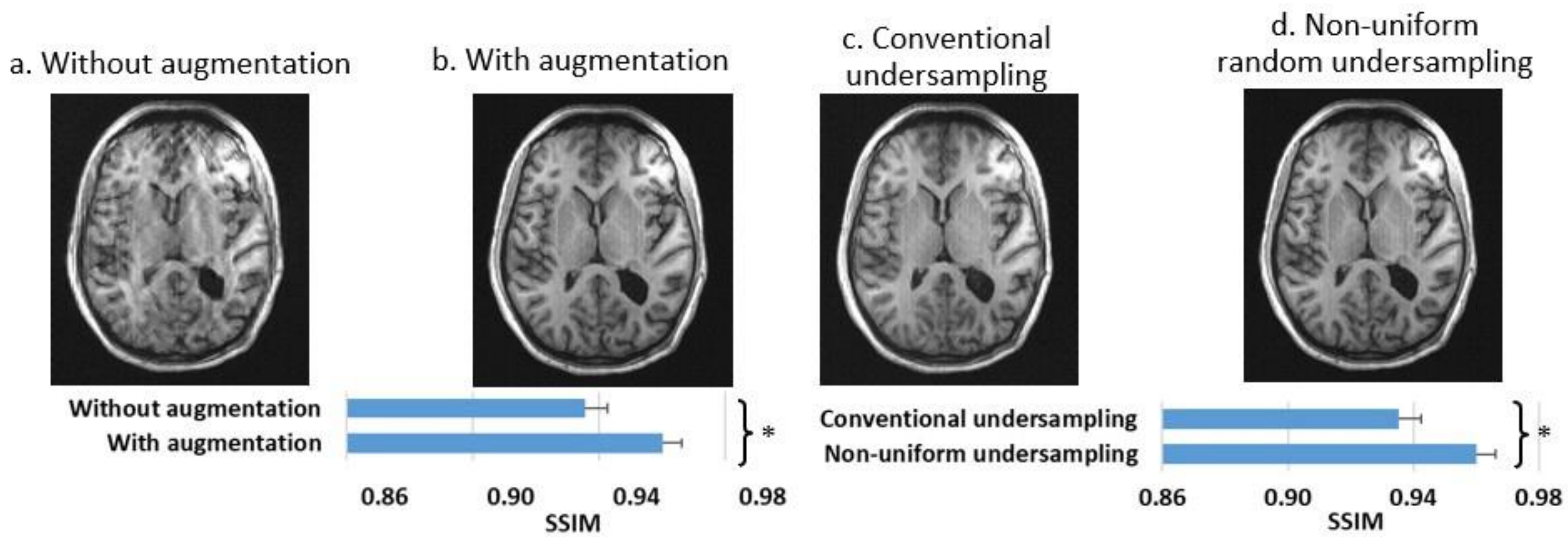

**Figure 9.** Comparison of reconstructions: a-b) with/without data augmentation c-d) with proposed and conventional undersampling

**Table 2.** DeepMRIRec: Reconstruction Outcomes

| RT-coils | SSIM | PSNR | Dice on HFC | HD on HFC |
|---|---|---|---|---|
| Compressed coil: 2 | 0.9646 | 28 | 0.66 | 72 |
| Compressed coil: 3 | 0.9555 | 27 | 0.66 | 71 |
| Compressed coil: 4 | 0.9557 | 26 | 0.64 | 70 |
| Loop coil | 0.9588 | 28 | 0.65 | 72 |
| 12 coils | 0.9556 | 26 | 0.64 | 71 |

## 5 CONCLUSION

In this manuscript, we presented DeepMRIRec for RT-coil specific MRI reconstruction. DeepMRIRec outperforms state-of-the-art methods, further closing the gap needed to meet demands of mission critical applications. This study shows that MRI acquisition can be accelerated by four-fold using DeepMRIRec with minimal compromise in internal details. The method has the potential to reduce the therapeutic burden carried by patients and to speed up both therapeutic planning and diagnosis.

## DATA AVAILABILITY STATEMENT

Source code, model weights and sample images are available at `https://github.com/stjude/DeepMRIRec`. The dataset will be provided based upon request with institutional approval.

## 6 ETHICS STATEMENT

This study involving human imaging data was approved by St. Jude Institutional Review Board (IRB Number: 22-1257) and conducted in accordance with the local legislation and institutional requirements. Written informed consent for participation in this retrospective study was not required from the participants or the minor(s)' legal guardian/next of kin in accordance with the national legislation and institutional requirements.

## AUTHOR CONTRIBUTIONS

SA: conceptualization, methodology, DeepMRIRec pipeline development, validation, writing original draft and visualization. JU and AD: data collection, data curation, evaluation of contemporary methods, and radiation oncology expert validation. CH: administration, supervision and editing. KK: conceptualization, writing original draft, and supervision. All authors contributed to the article and approved the submitted version.

## 7 FUNDING

This work was supported in part by the American Lebanese Syrian Associated Charities (ALSAC) and the National Cancer Institute (NCI Cancer Support Grants P30CA021765).

## ACKNOWLEDGMENTS

We wish to thank Mia Panlilio for early discussions in this project.

## 8 SUPPLEMENTARY MATERIAL

## 9 CONFLICT OF INTEREST

The authors have no conflict of interest to report.